\documentclass[fleqn,10pt, preprint, showpacs, superscriptaddress, prX]{revtex4-1}

\pdfoutput=1
\usepackage{amsmath}
\usepackage{graphicx}
\usepackage{caption}
\usepackage{subcaption}

\begin{document}
\makeatletter

\title{Universal fluctuations in growth dynamics of economic systems}
\author{Nathan C. Frey}
\affiliation{Department of Physics, Boston University, Boston, MA 02215, USA}
\author{Sakib Matin}
\affiliation{Department of Physics, Boston University, Boston, MA 02215, USA}
\author{H. Eugene Stanley}
\affiliation{Department of Physics, Boston University, Boston, MA 02215, USA}
\affiliation{Center for Polymer Studies, Boston University, Boston, MA 02215, USA}
\author{Michael Salinger}
\affiliation{Department of Markets, Public Policy and Law, Questrom School of Business, Boston University, Boston, MA 02215, USA}

\begin{abstract}
The growth of business firms is an example of a system of complex interacting units that resembles complex interacting systems in nature such as earthquakes. Remarkably, work in econophysics has provided evidence that the statistical properties of the growth of business firms follow the same sorts of power laws that characterize physical systems near their critical points. Given how economies change over time, whether these statistical properties are persistent, robust, and universal like those of physical systems remains an open question. Here, we show that the scaling properties of firm growth previously demonstrated for publicly-traded U.S. manufacturing firms from 1974 to 1993 apply to the same sorts of firms from 1993 to 2015, to firms in other broad sectors (such as materials), and to firms in new sectors (such as Internet services). We measure virtually the same scaling exponent for manufacturing for the 1993 to 2015 period as for the 1974 to 1993 period and virtually the same scaling exponent for other sectors as for manufacturing. Furthermore, we show that fluctuations of the growth rate for new industries self-organize into a power law over relatively short time scales.
\end{abstract}

\maketitle

\section*{Introduction}

Recently, the pursuit of statistical regularities in economics data and theoretical explanations have received increasing interest from both the physics and economics communities \cite{serino2011,gabaix2016}. Using data on US manufacturing firms from 1974 to 1993, Stanley et al. \cite{stanley1996,nunes1997} documented that the standard deviations of the growth rates obey a power law with a scaling exponent of approximately -1/5 \cite{nunes1997,buldyrev1997,stanley2002,riccaboni2008} and that and that the distribution of growth rates conditional on initial size is exponential over seven orders of magnitude \cite{stanley1996,nunes1997}. These results resemble the power laws that are robust statistical properties of many complex interacting physical systems \cite{gabaix2016,newman2007,reed2001,plerou2006,gabaix2003}. The theoretical explanation for such findings remains unclear. Models of critical phenomena in systems of strongly interacting elements predict results like those that have been found for firm growth, but so do models of weakly or non-interacting units \cite{gabaix2009, gabaix2016,amaral1998,amaral2001,lee1998}. Yet, to the extent that the above-mentioned results about the statistical properties of firm growth are stable, distinguishing among the competing explanations could provide important insights into the fundamental economic question of the nature of business firms \cite{buldyrev2016}. On the other hand, if changes in economic conditions cause the statistical properties of firm growth to change significantly, then the need for a theoretical explanation is less compelling. Therefore, a natural question to ask about empirical relationships in economics is whether they are as stable as power laws in physical systems or, alternatively, whether they change or even fall apart as the economy changes.

\section*{Results}

Fig. 1a is a log-log plot of the standard deviation of one-year growth rates as a function of initial firm size for U.S. manufacturing firms over two time periods. Firm growth rate is defined as $R = S_{1} / S_{0} $, where $S_{1}$ and $S_{0}$ are consecutive annual measures of firm size. The two time periods are 1974-1993 (“Original”), which is the period Stanley et al. analyzed, and 1993-2015 (“New”). The standard deviation, $\sigma(S_{0})$, is fit to a power law of the form $\sigma(S_{0}) = aS_{0}^{-\beta}$, where $a$ is a constant and $\beta$ is the scaling exponent. For the Original time period $\beta=0.25\pm.02$. For the New time period $\beta=0.26\pm0.01$. They are virtually identical to each other, and the results for the Original time period are not statistically significantly different from the 0.20 $\pm$ 0.03 reported by Stanley et al. for the same time period \cite{nunes1997}. This agreement suggests that the power law, initially proposed and verified in 1996 \cite{stanley1996}, is quite robust.

\begin{figure}[ht]
\centering
\includegraphics[width=\linewidth]{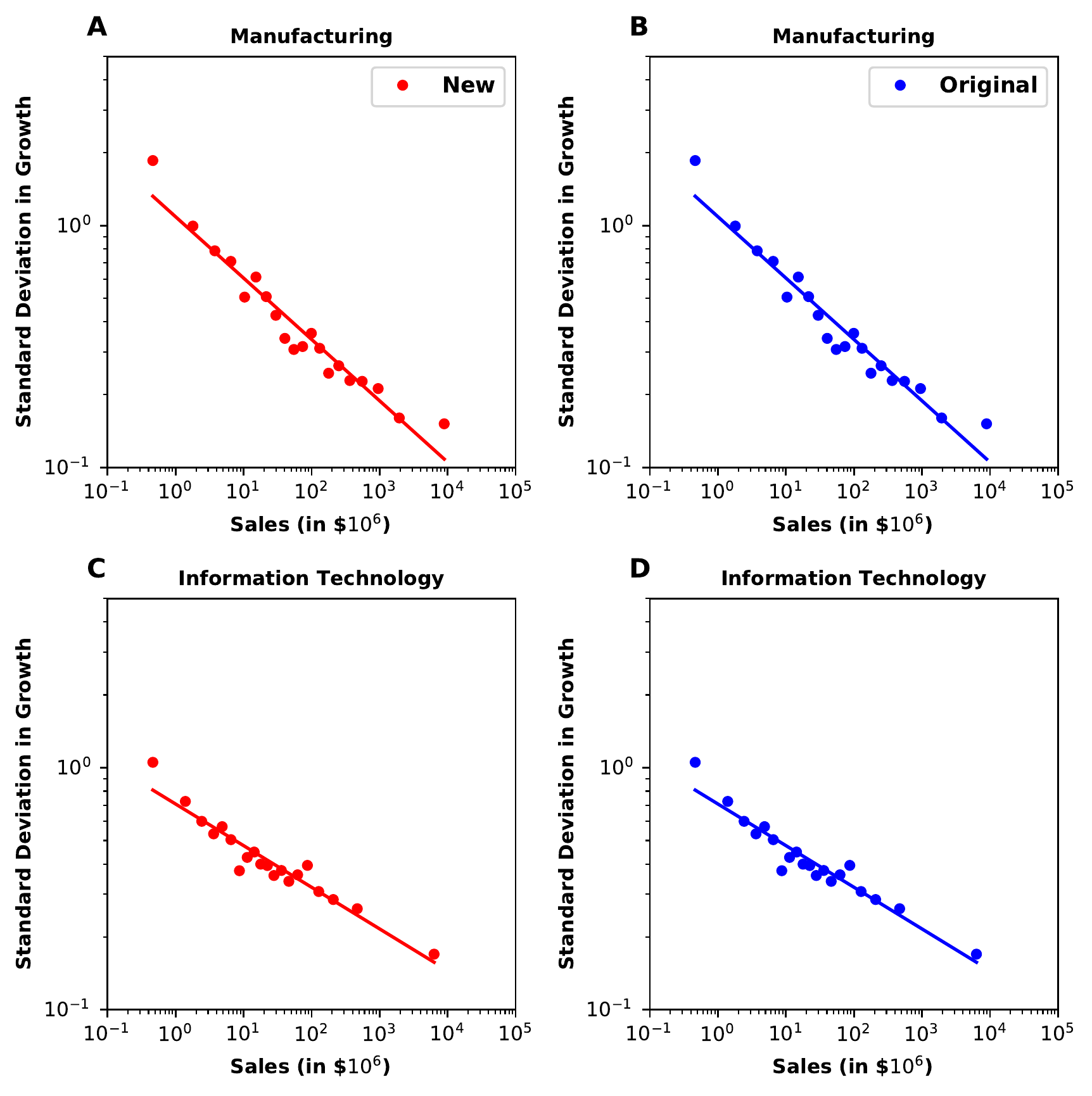}
\caption{Scaling of fluctuations against growth for 'Manufacturing' and 'Information Technology' in 'Original' and 'New' time periods. The stability of the exponent over all time is strong evidence of universality.}
\label{Fig:Pooled}
\end{figure}

We then extend our analysis to the ten other sectors (as categorized under the Global Industrial Classification System). For eight of the ten, we find a power law with the same (within error bars) scaling exponent as we measured for manufacturing. For example, for the 1974-1993 datasets, we find $\beta=0.20\pm 0.02,0.22\pm0.02,0.25\pm0.02$, and $0.22 \pm0.02$ for the financial, industrial, materials, and information technology sectors, respectively. Similarly, for more recent data from 1993-present, we find $\beta=0.18\pm 0.06,0.25\pm 0.01,0.23\pm 0.02$, and $0.25\pm 0.01$ for the financial, industrial, materials, and information technology sectors, respectively. Not only do fluctuations of the firm growth for each of these sectors obey a well-defined power law, the scaling exponents are approximately the same value for each sector. Plots for other sectors can be found in Fig. S1. Scaling exponents for all sectors over the Original and New time periods are reported in Table S1 and Table S2, respectively. Interestingly, utilities and finance are notable exceptions to the generality of the power law scaling.

\begin{figure}[ht]
\centering
\includegraphics[width=\linewidth]{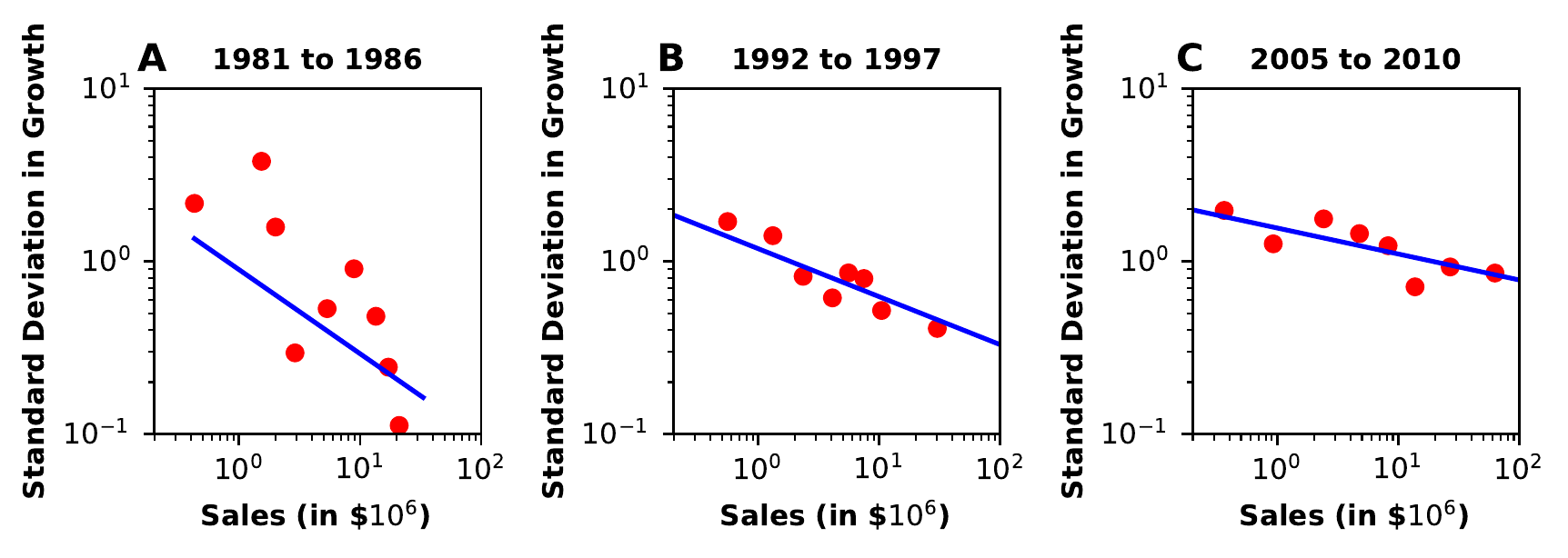}
\caption{Scaling of Biotechnology Industry over 3 distinct time periods showing self-organization of a power law.}
\label{Fig:BioT_SelfOrg}
\end{figure}

An interesting question to consider is whether the growth dynamics of new sectors in the economy (such as biotechnology) have the same properties as those of more established sectors. If the statistical properties of firm growth in fact are an example of self-organized criticality, we might expect the scaling properties associated with critical phenomenon to emerge over time. To test this hypothesis, we identified three industries for which we have data from their inception: ‘Biotechnology,’ ‘Software,’ and “Internet Software and Services.’ As above, we computed the standard deviation of logarithmic growth rates conditional on initial size for moving 5-year windows. For each 5-year window, we regressed the logarithm of the standard deviation on the logarithm of initial size. Fig. \ref{Fig:BioT_SelfOrg} shows the results for the ‘Biotechnology’ sector. In Fig. 2a, which shows results for the window from 1981 to 1986, the SE is 0.2 and it is clear that there is no power law relationship between the data. In contrast, over the time periods 1992-1997 (Fig. 2b) and 2005-2010 (Fig. 2c), a power law trend clearly emerges. The regression SE reduces to 0.02, and the scaling law accurately describes standard deviation in growth over 4 orders of magnitude. At the “birth” of a sector, few firms exist and there is no evidence of organization. Over the time period underlying Fig. 2a, there were only 53 publicly traded American biotechnology firms included in the analysis. The number increases to 214 (Fig. 2b) and 514 (Fig. 2c). Plots showing similar self-organization behavior are shown for the recently established Software and Internet Software \& Services industries in Fig. S2 and Fig. S3, respectively. Future work may probe and explain the dynamics of this self-organization behavior.

\begin{figure}[ht]
\centering
\includegraphics{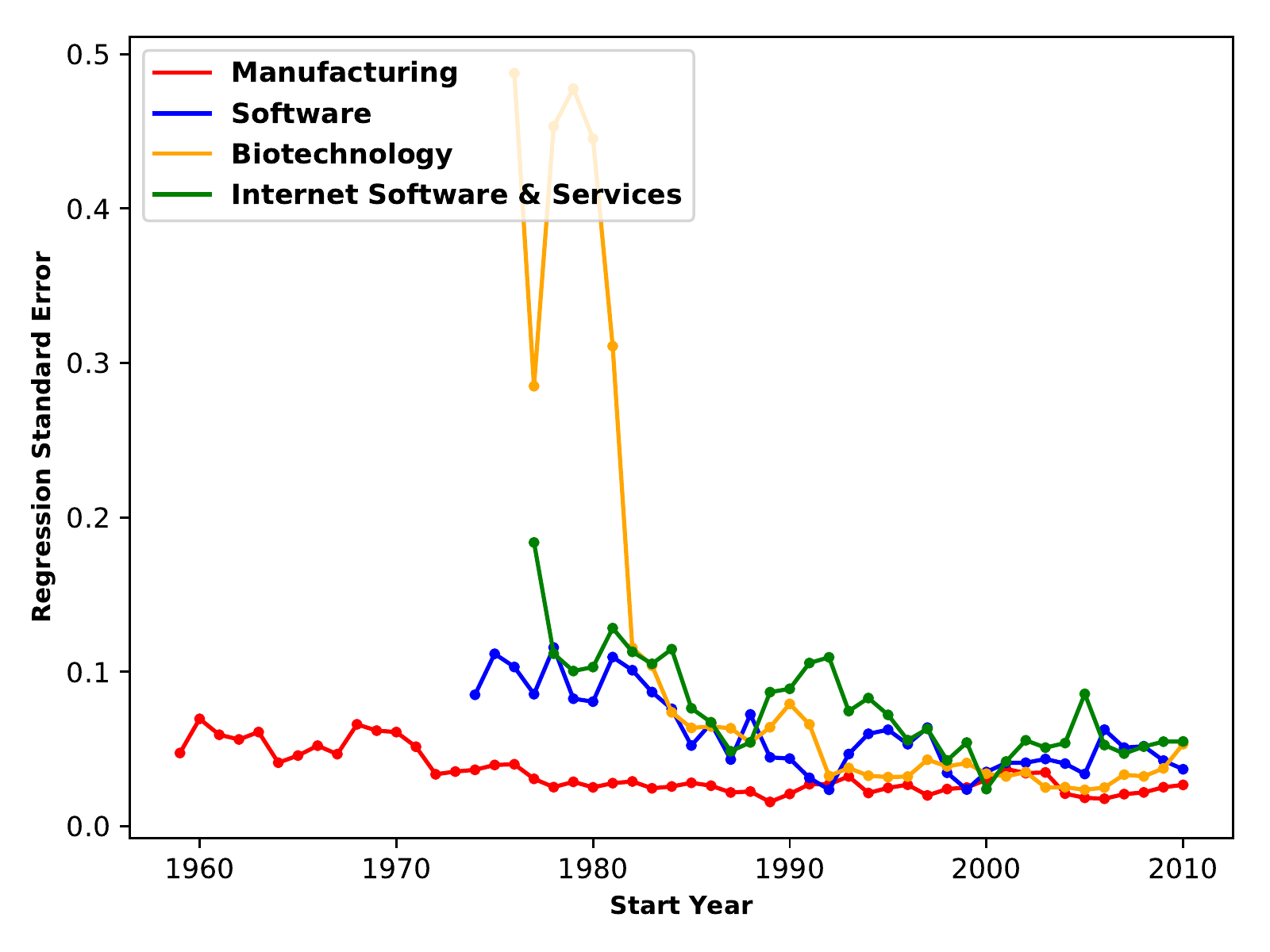}
\caption{Times series of the regression standard error for different industries showing fast self-organization of power laws.}
\label{Fig:SSE}
\end{figure}

As seen in Fig. 2, nascent sectors self-organize over time to conform to the same universal scaling relation seen in established sectors like manufacturing. We extend the above analysis by considering time series of regression standard errors (SE) for pooled data for a variety of sectors, including biotechnology. As shown in Fig. \ref{Fig:SSE} (red), the standard errors for the manufacturing sector regressions are fairly low, indicating good fit to a power law. Furthermore, we see there is a near-monotonic decrease of the SE.

We analyze the 5-year pooled regressions for newly established industries for which we have data from their inception: the 'Biotechnology', 'Software', and 'Internet Software and Services' industries.  As seen in Fig. \ref{Fig:SSE}, the time series of the SE for all three industries shows a sharp decrease over some characteristic time scale. The SE for Biotechnology in 1980 is 0.5 and decreases to below 0.1 from 1984 to the present year. Similar trends are observed if we use the R-squared goodness of fit metric instead of SE. From this straightforward analysis, we conclude that there is a "convergence" towards power law behavior which is universal across different types of companies. Possible "convergence criteria" and models have been proposed \cite{amaral1998,lee1998,amaral2001,bottazzi2006,digiovanni2011,fu,buldyrev2016}, which can be tested in future work. 

\section*{Conclusion}
The results presented above enhance our understanding of the empirical facts that describe the dynamics of firm size and growth. The robustness of the observed scaling laws across different sectors over many orders of magnitude provides compelling evidence that general dynamical principles, not specific to particular industries, govern the growth of firms. The analysis of ‘new industries’ illuminates growth dynamics at the early stages. The self-organization of the scaling behavior for new industries provides new, stronger evidence of universality in economic systems.

\section*{Methods}

Data was collected from the Compustat database, which is available through Wharton Research Data Services. Compustat data includes firms representing over 99\% of global market capitalization, from 1950 to the present. 

Companies were sorted according to the Global Industry Classification Standard (GICS). At the highest level, the economy is broken into “sectors." In specific cases, we show power law behavior from the sector level, down through the “industry group” and “industry” divisions.

Several metrics for firm size can be used, such as sales, employees, etc. In the paper, we used the 'net sales' or revenue as the firm size, as is standard in the economic literature. We have conducted a similar analysis for employees and assets and the results are qualitatively similar to that for sales. These results are summarized for the manufacturing sector in Fig. S4.

We analyze the standard deviation of growth rates, $R$ versus initial size, $S_{0}$. The log-log scale is chosen such that a straight-line plot on the log-log plot corresponds to a power law, and the slope of the line corresponds to the scaling exponent. We extract the exponent by running a regression on a log-log scale. For each plot, we pooled all the firms' initial sizes and yearly growth rates within a given industry or sector and over a given period of time. To compute the standard deviations for bins of growth rates, we used 20 bins when data is plotted at a sector level over the 'Original' and 'New' time period. The 'Original' time corresponds to the same years as \cite{stanley1996} whereas the 'New' time corresponds to all subsequent years. When analyzing our new industries, we used 10 bins. We computed scaling exponents for 10, 20, and 30 bins for the manufacturing sector to verify that the computed exponents do not depend on the number of bins. These plots are shown in Fig. S4 and the scaling exponents are reported in Table S3. One-year growth rates of greater than 1000\% were discarded as outliers. 

Stanley et al. analyzed the manufacturing industry between 1974 to 1993. The original data set was classified using the Standard Industrial Classification (SIC) method. The data available to us now follows the GICs classification. As such, the 'original data' was constructed by pooling together appropriate industries to reconstruct the 'manufacturing' sector that was analyzed in \cite{stanley1996}.

We analyzed the stability of the power law by using 5-year pooled regressions. For a given starting period, we pooled the yearly growth rates and initial sizes for all companies in that sector or industries for five subsequent years. This step is repeated by shifting the starting and end years by one. The data was pooled for five years as the year-to-year plots were noisy for smaller industries.

\section*{Supplementary Information}
\renewcommand{\thefigure}{S\arabic{figure}}
\renewcommand{\thetable}{S\arabic{table}}
\setcounter{figure}{0}

\begin{figure}[ht]
\centering
\includegraphics{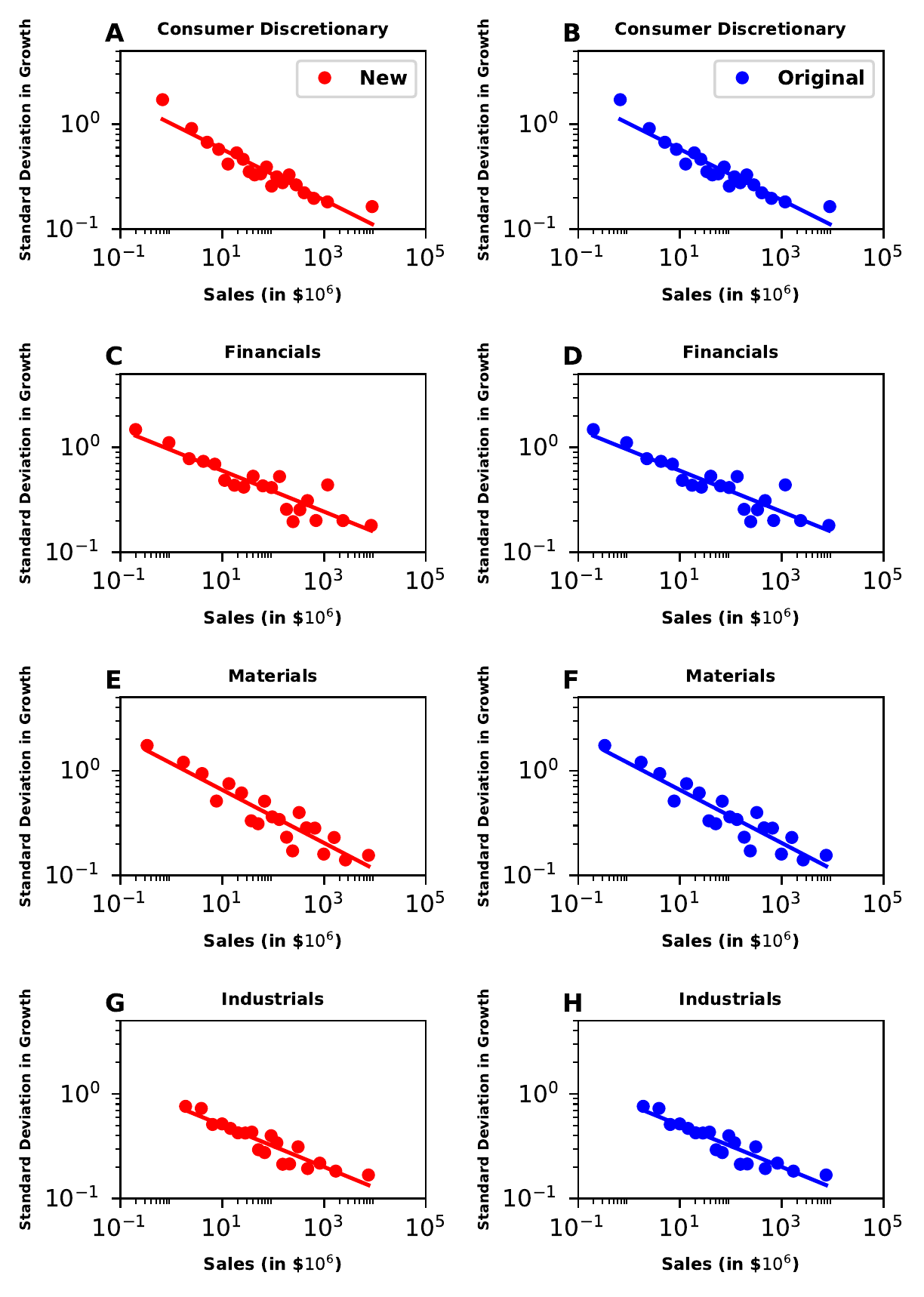}
\caption{Scaling of fluctuations against growth for 'Consumer Discretionary', 'Financials', 'Industrials', and 'Materials' in the 'New' (left) and 'Original' (right) time periods. The stability of the exponent in both time frames is strong evidence of universality.}
\label{Fig:Pooled_SI}
\end{figure}

\clearpage 

\begin{figure}[ht]
\centering
\includegraphics{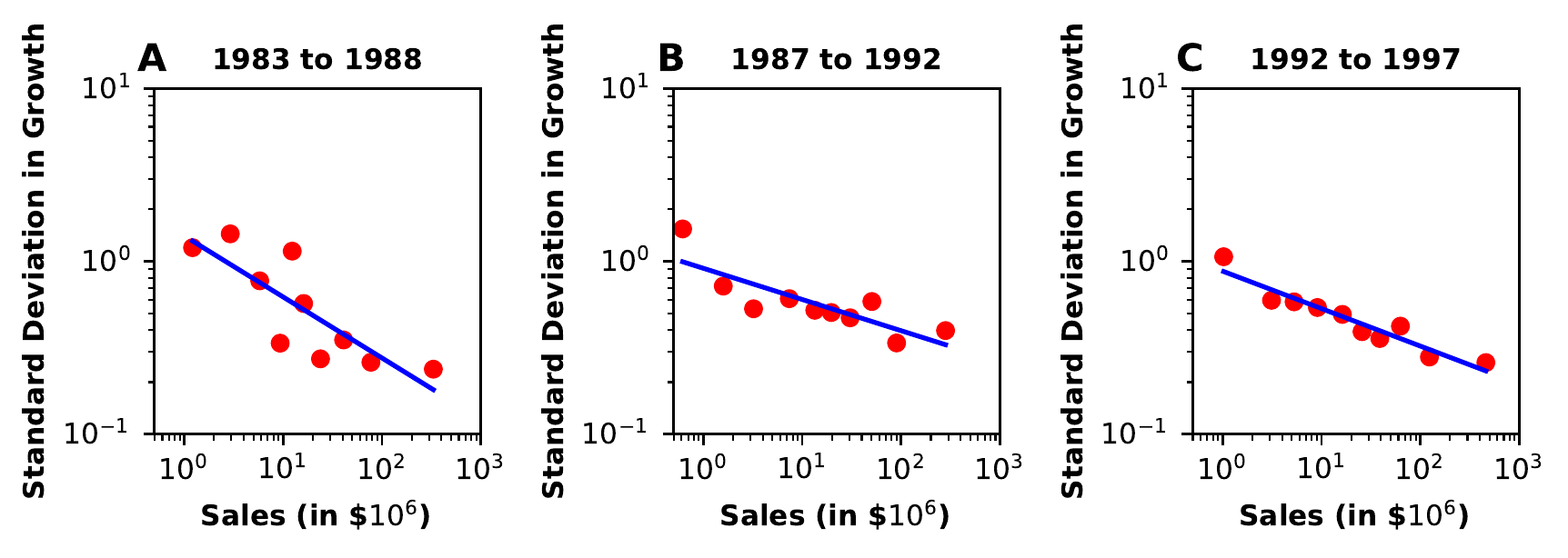}
\caption{Scaling of \textbf{Software} Industry at 3 distinct times showing self-organization of a power law.  }
\label{Fig:Software_SelfOrg}
\end{figure}

\begin{figure}[ht]
\centering
\includegraphics{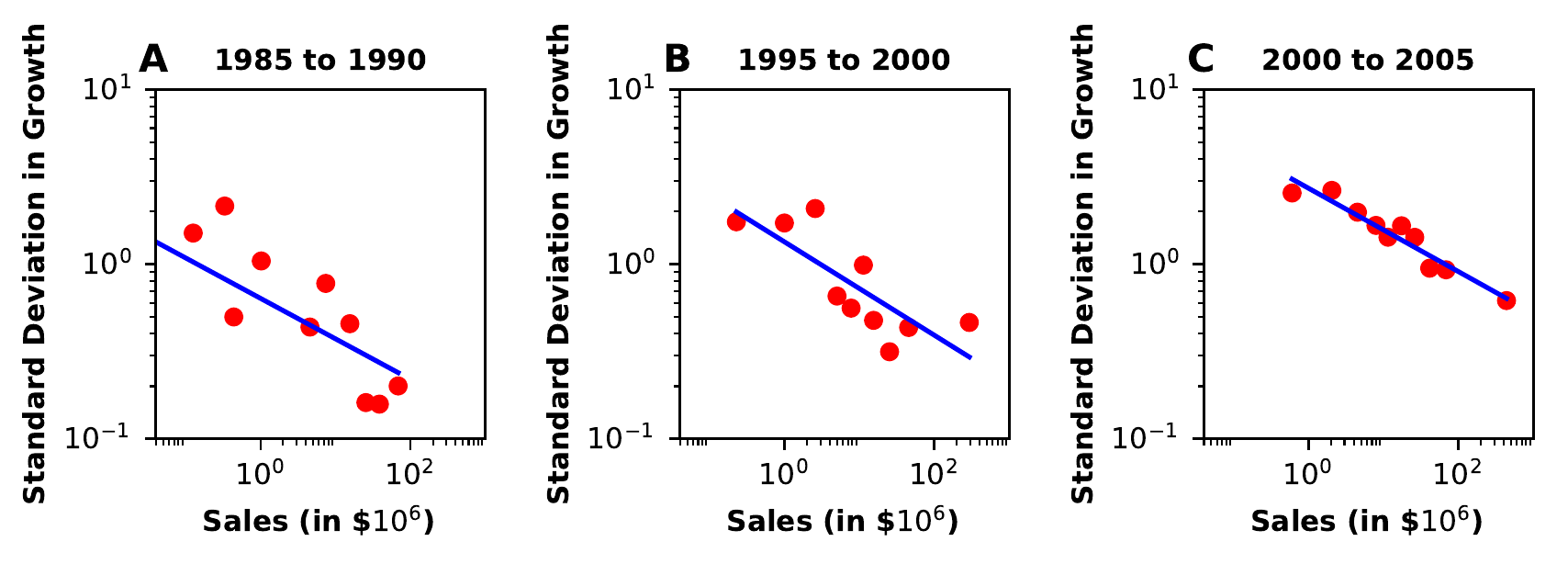}
\caption{Scaling of \textbf{Internet Software \& Services} at 3 distinct times showing self-organization of a power law.}
\label{Fig:Software_IntOrg}
\end{figure}

\clearpage 

\begin{figure}[ht]
\centering
\includegraphics{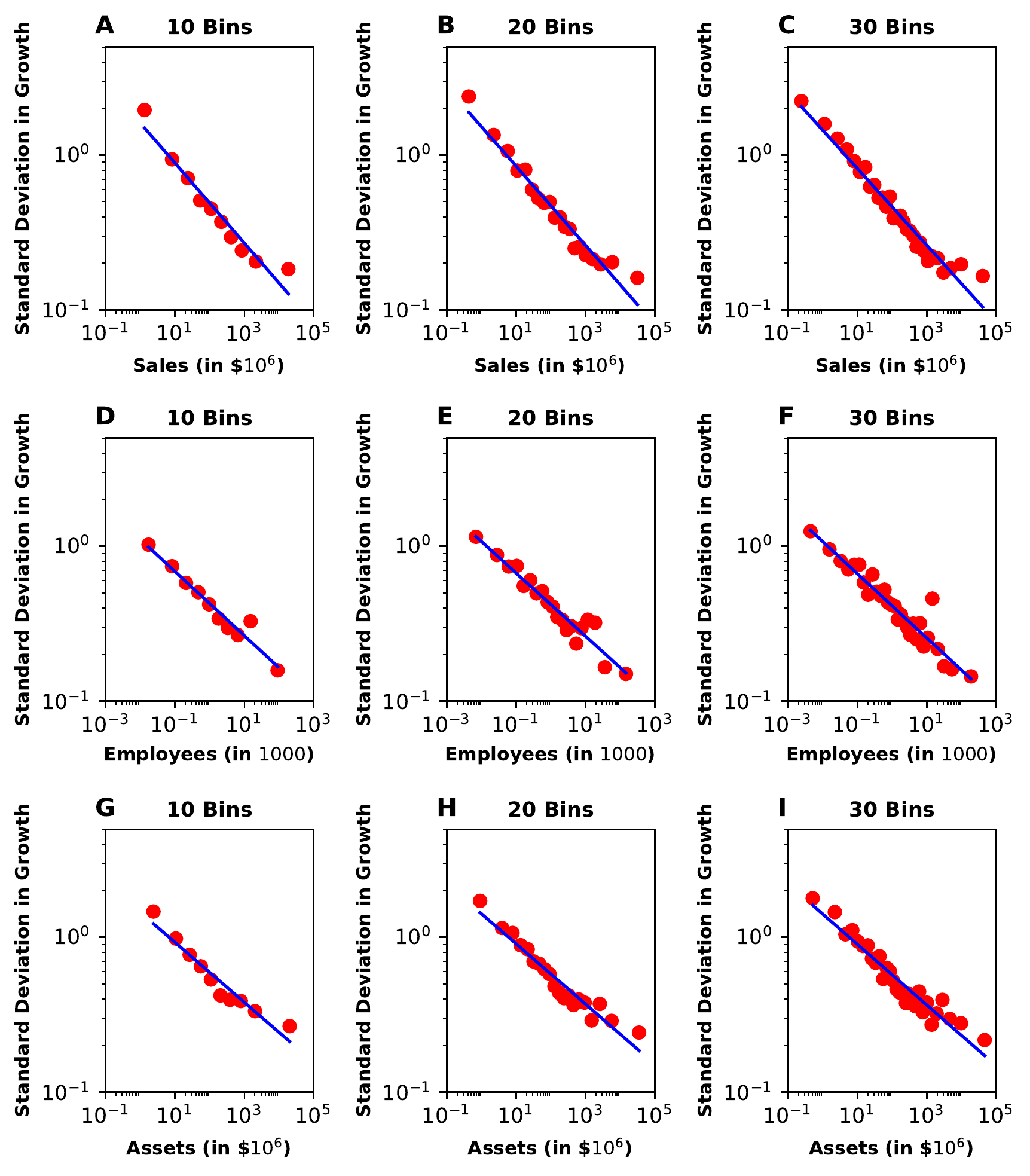}
\caption{Scaling of Manufacturing sector in the 'New' time period with sales, employees, and assets as measures of growth, plotted with 10, 20, and 30 bins. The value of the scaling exponent is the same (within error) regardless of the number of bins.}
\label{Fig:bin}
\end{figure}

\clearpage 

\begin{table}[htp]
\centering
\begin{tabular}{|l|l|l|l|l|l|}
Name                   & Slope \ \ \ \  & Intercept \ \ \    & RSqr \ \ \ \  & Std-Err \ \ \    & No.Data Points \\
Manufacturing         & -0.253 & 0.084 & 0.939 & 0.0152 & 25527          \\
Energy                 & -0.189 & -0.112 & 0.938 & 0.0115 & 3413           \\
Materials              & -0.254 & 0.172   & 0.879 & 0.0221  & 3807           \\
Industrials            & -0.220  & -0.127 & 0.851 & 0.0217  & 9372           \\
Consumer Discretionary & -0.243 & 0.011  & 0.895 & 0.0196 & 10505          \\
Consumer Staple        & -0.235   & -0.156 & 0.734 & 0.0333 & 2650           \\
Health Care            & -0.248  & 0.009  & 0.854 & 0.0248 & 3180           \\
Financials             & -0.195 & -0.059 & 0.824 & 0.0213 & 5728           \\
Information Technology & -0.224 & -0.060 & 0.849 & 0.0223 & 5533           \\
Telecommunication      & -0.290 & 0.200  & 0.656 & 0.0495  & 1182           \\
Utilities              & -0.242 & -0.288 & 0.680 & 0.0391  & 3086  
\end{tabular}
\caption{Summary of scaling of growth rates for all sectors classified under GICS during `Original' Time. The fluctuations of the growth rates as a function of the initial size decays as a power law for all industries.  }
\label{Tb_Nature}
\end{table}

\clearpage 

\begin{table}
\centering
\begin{tabular}{|l|l|l|l|l|l|}
Name                   & Slope  \ \ \ \ \   & Intercept \ \ \    & RSqr \ \ \ \   & Std-Err \ \ \  & No.Data Points \\
Manufacturing          & -0.256 & 0.421  & 0.957 & 0.0127 & 40908          \\
Energy                 & -0.189 & 0.465  & 0.938  & 0.0115 & 6792           \\
Materials              & -0.232 & 0.438  & 0.913 & 0.0170 & 5955           \\
Industrials            & -0.253 & 0.323  & 0.948 & 0.0140 & 11923          \\
Consumer Discretionary & -0.266 & 0.465  & 0.918 & 0.0190 & 13989          \\
Consumer Staples       & -0.215 & 0.062  & 0.853 & 0.0210 & 4094           \\
Health Care            & -0.245 & 0.401  & 0.958 & 0.0121 & 10081          \\
Financials             & -0.176 & 0.080  & 0.305  & 0.0626 & 16962          \\
Information Technology & -0.248 & 0.484  & 0.943 & 0.0143 & 16097          \\
Telecommunications     & -0.257 & 0.832  & 0.908 & 0.0193 & 2087           \\
Utilities              & -0.105 & -0.602  & 0.193 & 0.0508 & 3318          
\end{tabular}
\caption{Summary of scaling of growth rates for all sectors classified under GICS during 'New' Time. As a function of the initial size, the fluctuations in the growth rates decrease as a power law, for all sectors except Utilities. }
\label{Tb_New}
\end{table}

\clearpage 

\begin{table}[htp]
\centering
\begin{tabular}{|l|l|l|l|l|l|l|}
Measure   \ \ \    & Bins \ \ \           & Slope \ \ \ \  & Intercept \ \ \    & RSqr \ \ \ \  & Std-Err \\    
Sales & 10 & -0.194 & 0.37 & 0.04 & 0.018 \\
Sales & 20 & -0.195 & 0.35 & 0.94 & 0.012 \\
Sales & 30  & -0.196 & 0.35 & 0.93 & 0.010  \\
Employees & 10 & -0.208 & 0.85 & 0.95 & 0.017 \\
Employees & 20 & -0.204 & 0.87 & 0.93 & 0.013 \\
Employees & 30  & -0.208 & 0.89 & 0.91 & 0.012  \\
Assets & 10 & -0.259 & 0.48 & 0.94 & 0.022 \\
Assets & 20 & -0.256 & 0.42 & 0.96 & 0.013 \\
Assets & 30  & -0.248 & 0.37 & 0.96 & 0.010  \\  
\end{tabular}
\caption{Summary of scaling of growth rates for the Manufacturing sector classified under GICS during `Original' Time, for different measures of growth and using different number of bins.  }
\label{Tb_Bin}
\end{table}

\end{document}